\documentclass[slac_one]{revtex4}
\usepackage{graphicx}
\usepackage{fancyhdr}
\begin{document}

\title{Testing the Standard Model with $W\gamma$ and $Z\gamma$ at the Tevatron} %% Paper title goes here


\author{Adam L. Lyon (for the CDF and D0 collaborations)}
\affiliation{Fermi National Accelerator Laboratory, Batavia, IL 60510, USA}

\begin{abstract}
  Results on analyses involving $W\gamma$ and $Z\gamma$ production
  from the CDF and D0 experiments at the Fermilab Tevatron Collider
  at $\sqrt{s} = 1.96$~TeV are presented here. Using 1-2~fb$^{-1}$ of
  data, cross sections, anomalous coupling limits, and the $W\gamma$
  Radiation Amplitude Zero are reviewed.
\end{abstract}

\maketitle

\thispagestyle{fancy}


\section{INTRODUCTION} % Section title should be in all capitals.

Boson self-interactions are a consequence of $SU(2)_L \times U(1)_Y$
gauge symmetry. Certain self-interactions such as $WW\gamma$ are
required by the standard model (SM), while others such as $ZZ\gamma$
and $Z\gamma\gamma$ are forbidden. If these trilinear-gauge couplings
are not as the SM predicts, then new physics may lead to striking
deviations in the kinematics of the process compared to the SM
prediction, such as enhancement of the photon energy spectrum at
large $E_T^{\gamma}$.

The $W\gamma$ process offers another window on new physics. Gauge
theory predicts that any four particle tree amplitude involving one or
more gauge bosons may be factorized into a part depending on charge
alone and a second part depending on spin and polarization. At a
particular point in phase space, the charge part will cause the
amplitude to vanish. For $W\gamma$, this effect leads to a
zero\cite{radzero} (the Radiation Amplitude Zero, RAZ) in the angular
distribution of the photon in the CM frame of the incoming quarks. But
in hadron collisions determination of this frame is difficult because
the direction of the $\nu$ that decayed from the $W$ is
unknown. Therefore instead, the charge-signed rapidity difference\cite{CSRD}
defined to be $Q_\ell \times \Delta y$ ($Q_\ell$ is the charge of the
$\ell$ decayed from the $W$ and $\Delta y$ is the $\ell-\gamma$
rapidity difference) is used. At the Tevatron, the psuedorapidity
difference ($\Delta \eta$) is an accurate replacement for $\Delta
y$. The RAZ then appears as a dip in this distribution at $Q_\ell
\times \Delta \eta \approx -1/3$. Non-SM trilinear-gauge couplings act
to wash out the dip as does final state radiation.

The key to the analyses described here is excellent photon
identification. Photons appear similar to electrons in detectors, but
photons do not have an associated track. Both CDF and D0 utilize
preshowers to enhance discrimination of photons against jets. D0 uses
calorimeter layers as well as the preshower measurement to point back
to the vertex of the photon to further reject jets. CDF also removes
photons generated by electron bremsstrahlung with tracks in the
silicon detector. Photon efficiency is mainly determined by Monte
Carlo studies, but for low energies it is possible to verify the
simulations with low energy final state radiation $Z\gamma$ events.

\section{$Z\gamma \rightarrow \ell \ell \gamma$}

\begin{figure}
\includegraphics[height=0.9in]{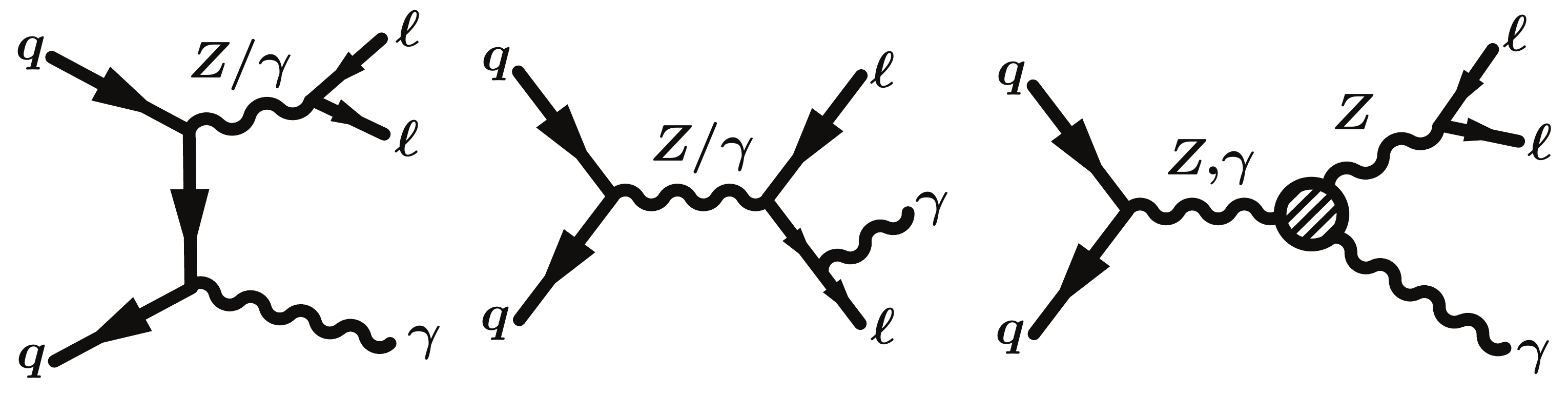}%
\caption{\label{fig:zgam_feyn} Diagrams for $Z\gamma$ production: ISR, FSR
and the forbidden $s$-channel involving the trilinear gauge coupling.}
\end{figure}

In the SM at tree level, $Z\gamma$ is produced by initial state
radiation (ISR) and final state radiation (FSR) diagrams shown in
Fig.~\ref{fig:zgam_feyn}. The SM forbids the $s$-channel diagram involving the
trilinear gauge couplings. Cross sections are
measured with $ee\gamma$ and $\mu\mu\gamma$ events. The D0
analysis\cite{d0zgam} is based on 1.1~fb$^{-1}$ of data while the CDF
analysis\cite{cdfzgam} is based on 1.1~fb$^{-1}$ of data in the
electron channel and 2~fb$^{-1}$ in the muon channel. Both analyses
require central photons with $E_T^{\gamma} > 7$~GeV and a minimum
$\Delta R$ between a lepton and the photon of 0.7. The minimum
di-lepton invariant mass is 30~GeV/$c^2$ and 40~GeV/$c^2$ for D0 and
CDF respectively.

\begin{figure}
\includegraphics[height=2.3in]{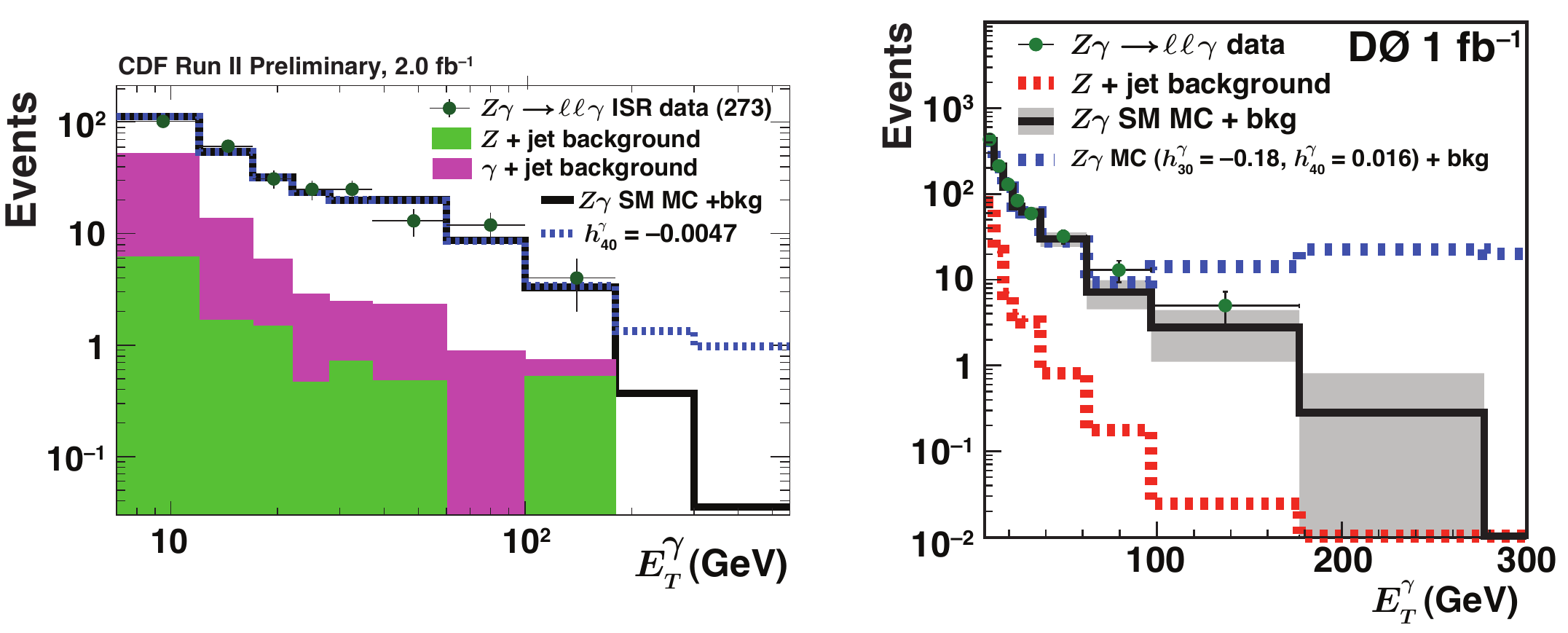}%
\caption{\label{fig:zgamet} Photon $E_T$ spectra in $Z\gamma$
  production for the CDF and D0 analyses respectively.}
\end{figure}

The cross section measurements are
shown in Table~\ref{tab:zgamxs}. The $E_T^{\gamma}$ spectra are shown
in Fig.~\ref{fig:zgamet}. No deviation from the SM prediction is
observed. 

\begin{table}
\caption{\label{tab:zgamxs}Cross section results for $Z\gamma$ analyses.}
\begin{tabular}{llrll}
 Value         &  Observed                                                          &  SM Expectation    &     &     \\
\hline
{\bf CDF}  &  $E_T^\gamma >$7 GeV, $\Delta R>$0.7, $M_{\ell\ell}>$40 GeV/$c^2$  &                    &     &     \\
 ISR+FSR       &  778 events (390 $ee\gamma$ + 388 $\mu\mu\gamma$)                  &  771 $\pm$ 41      &     &     \\
 ISR+FSR       &  $\sigma$(pb) = 4.6 $\pm$ 0.2(stat) $\pm$ 0.3(sys) $\pm$ 0.3(lum)  &  4.5 $\pm$ 0.4     &     &     \\
 ISR           &  $\sigma$(pb) = 1.2 $\pm$ 0.1(stat) $\pm$ 0.2(sys) $\pm$ 0.1(lum)  &  1.2 $\pm$ 0.1     &     &     \\
 FSR           &  $\sigma$(pb) = 3.4 $\pm$ 0.2(stat) $\pm$ 0.2(sys) $\pm$ 0.2(lum)  &  3.3 $\pm$ 0.3     &     &     \\
\hline
{\bf D0}  &  $E_T^\gamma >$7 GeV, $\Delta R>$0.7, $M_{\ell\ell}>$30 GeV/$c^2$  &                    &     &     \\
 ISR+FSR       &  968 events (453 $ee\gamma$ + 515 $\mu\mu\gamma$)                  &  920.4 $\pm$ 53.4  &     &     \\
 ISR+FSR       &  $\sigma$(pb) = 5.0 $\pm$ 0.3(stat+sys) $\pm$ 0.3(lum)             &  4.74 $\pm$ 0.22   &     &     \\
\end{tabular}
\end{table}

Limits are placed on on anomalous $ZZ\gamma$ and $Z\gamma\gamma$
couplings. $ZV\gamma$ couplings (where $V$ is either $Z$ or $\gamma$)
are parameterized by CP violating parameters ($h_1^V$, $h_2^V$) and CP
conserving parameters ($h_3^V$, $h_4^V$). All vanish in the SM. In the
case of non-SM physics, these couplings may rise with CM energy and
violate unitarity. To avoid that problem, it is typical to transform
these couplings with a form factor $h \rightarrow
h/(1+\hat{s}/\Lambda^2)^n$. Both experiments choose $\Lambda =
1.2$~TeV. 95\% CL limits on the real parts of the CP conserving
couplings are shown in Table~\ref{tab:zgamac}. For the Tevatron, lmits
on the CP violating couplings are nearly identical to the
corresponding CP conserving couplings. The combined LEP2
results\cite{lep2} are shown as well.

\begin{table}
  \caption{ \label{tab:zgamac} 95\% CL limits on $ZZ\gamma$ and
    $Z\gamma\gamma$ couplings.}

\begin{tabular}{cccc}
               &  CDF (1.1 fb$^{-1}$ $e$, 2.0 fb$^{-1}$ $\mu$)  &  D\O\ 1.1 fb$^{-1}$  &  LEP2 2003        \\
\hline
 $h_3^\gamma$  &  $[-0.084, 0.084]$                             &  $[-0.085,0.084]$    &  $[-0.049, -0.008]$  \\
 $h_4^\gamma$  &  $[-0.0047, 0.0047]$                           &  $[-0.0053,0.0054]$  &  $[-0.002, -0.034]$  \\
 $h_3^Z$       &  $[-0.083, 0.083]$                             &  $[-0.083,0.082]$    &  $[-0.20, 0.07]$     \\
 $h_4^Z$       &  $[-0.0047, 0.0047]$                           &  $[-0.0053,0.0054]$  &  $[-0.05, 0.12]$     \\
\end{tabular}
\end{table}

\section{$W\gamma \rightarrow \ell \nu \gamma$}

\begin{figure}
\includegraphics[height=0.9in]{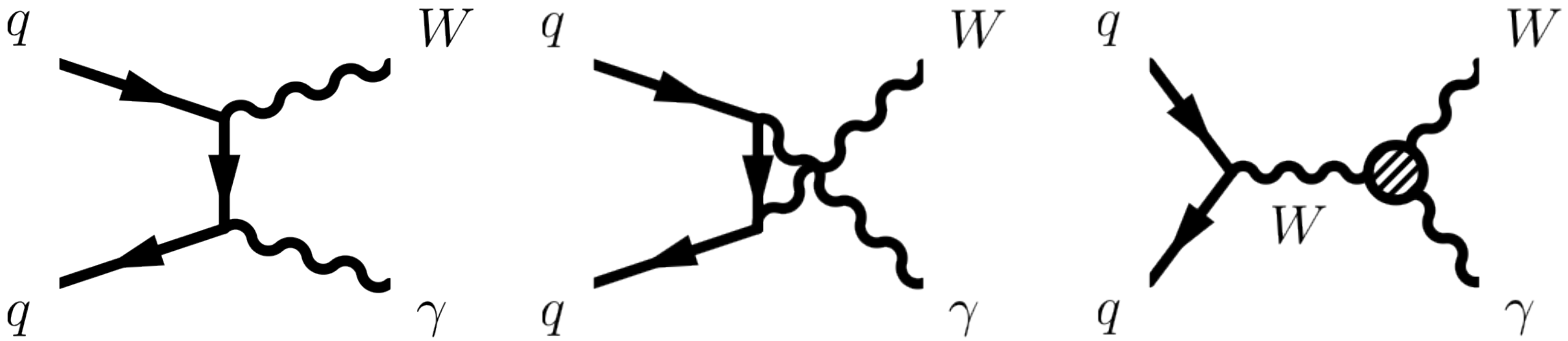}%
\caption{\label{fig:wgamdiag} ISR and $s$-channel diagrams for
  $W\gamma$ production. The FSR diagram is not shown.}
\end{figure}

\begin{figure}
\includegraphics[height=2.6in]{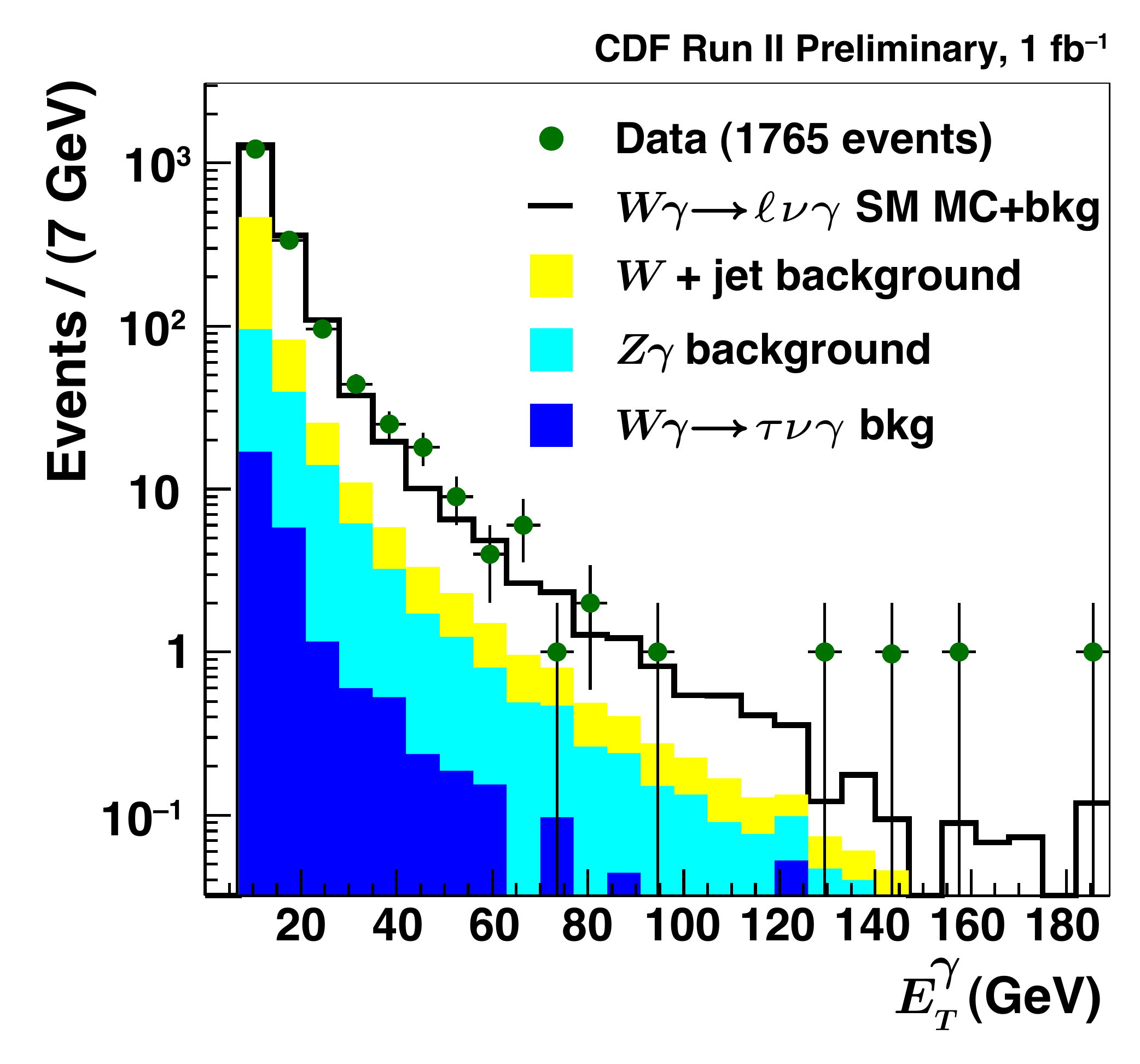}
\includegraphics[height=2.0in]{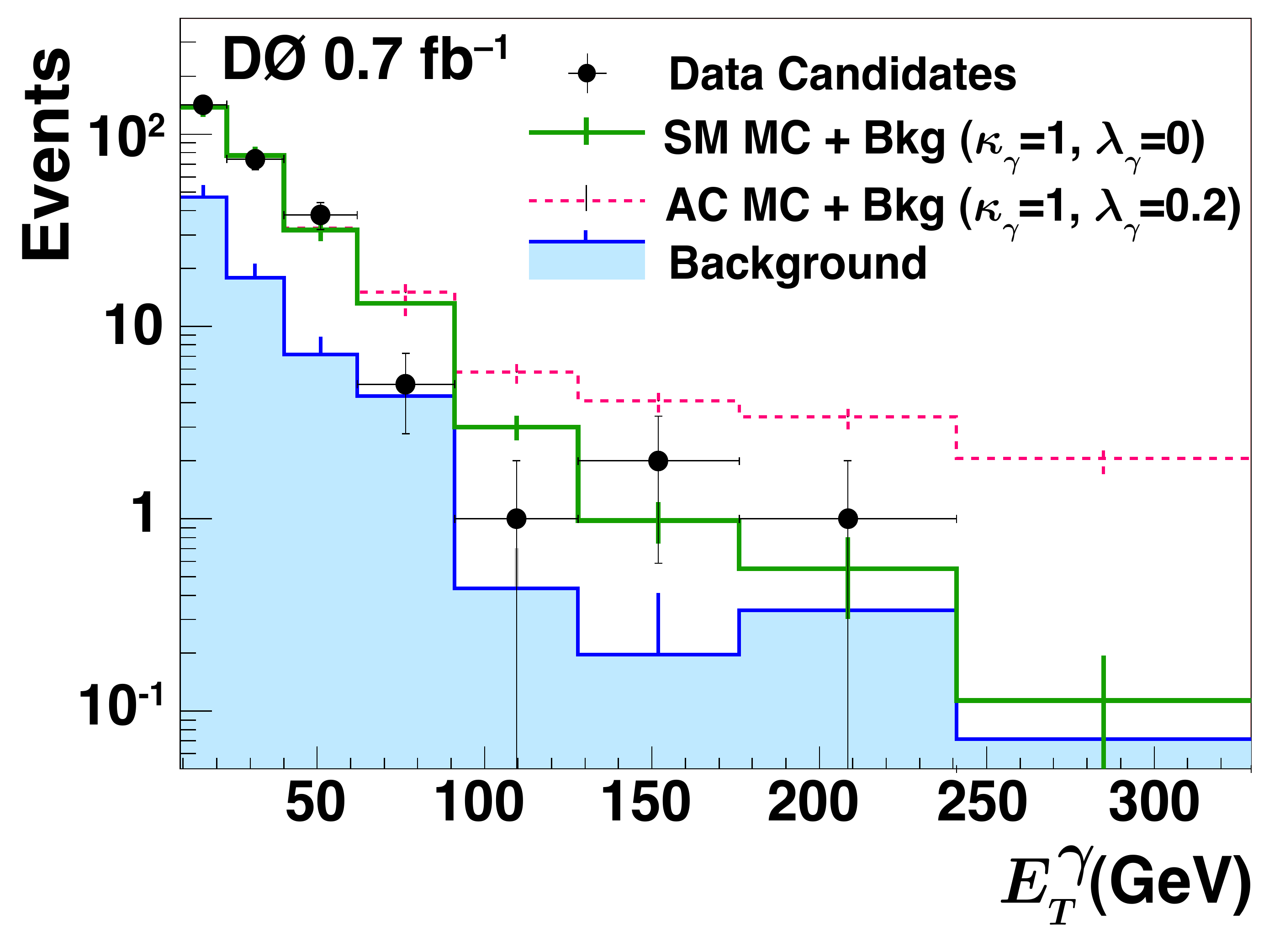}%%
\caption{\label{fig:phoet} Photon $E_T$ spectra for the CDF
 and D0 $W\gamma$ analyses respectively.}
\end{figure}

$W\gamma$ production proceeds through three diagrams plus FSR (the
$s$-channel diagram with the $WW\gamma$ coupling is required by the
SM) as shown in Fig.~\ref{fig:wgamdiag}. A CDF analysis\cite{cdfwgam}
measures the $W\gamma$ production cross section with 1~fb$^{-1}$ of
data with $W\gamma \rightarrow \ell\nu\gamma$ events where $\ell$ is
an $e$ or $\mu$. The analysis requires central photons with
$E_T^\gamma > 7$~GeV, missing $E_T > 25$ ($20$)~GeV for the $e$
($\mu$) analysis, $\Delta R_{\ell\gamma} > 0.7$, $\ell\nu$ transverse
mass between 30 and 120~GeV/$c^2$, and central electrons and muons
with $E_T^e > 25$~GeV and $E_T^\mu > 20$~GeV. The combined cross
section is measured to be $\sigma(W\gamma) \times
\text{BR}(W\rightarrow \ell\nu) = 18.03 \pm 0.65_{\text{stat}} \pm
2.55_{\text{sys}} \pm 1.05_{\text{lum}}$~pb. This value compares well
with the SM prediction of $19.3 \pm 1.4$~pb. The $E_T^\gamma$ spectrum
is shown in Fig.~\ref{fig:phoet}.

The corresponding D0 analysis\cite{d0wgam} is optimized to examine the
RAZ discussed above and examined 700~pb$^{-1}$ of data. Here, forward
photons are crucual for measuring $Q_\ell \times \Delta
\eta$. Requirements are central and forward photons with $E_T^\gamma >
9$~GeV, missing $E_T > 25 \, (20)$~GeV for the electron and muon analysis
respectively and $\Delta R_{\ell\gamma}  0.7$. The electron analysis
additionally requires a central or forward electron with $E_T >
25$~GeV, $e\nu$ transverse mass $> 50$~GeV/$c^{2}$, and three body
transverse mass $M_{T3}$ ($e$, $\gamma$, and missing $E_T$) $>
120$~GeV/c$^{2}$. The muon analysis requires a muons within $|\eta| <
1.6$ with $p_T > 20$~GeV/$c$ and $M_{T3} > 110$~GeV/$c^2$. The
$M_{T3}$ cuts are optimized to reject FSR events as they will obscure
the dip. The background subtracted signal yields are $130 \pm 14 \pm
3.4$ for $e\nu\gamma$ events and $57 \pm 8.8 \pm 1.8$ for the
$\mu\nu\gamma$ events (uncertainties are statistical and systematic
respectively). These results compare well to the SM predictions of
$120\pm 12$ and $77 \pm 9.4$ for $e\nu\gamma$ and $\mu\nu\gamma$
respectively. The measured $E_T^\gamma$ spectrum is displayed in
Fig.~\ref{fig:phoet}. No deviation from the SM is observed. Using the
photon $E_T$ spectrum, 95\% CL limits on anomalous $WW\gamma$
couplings are determined to be $0.49 < \kappa_\gamma < 1.51$, $-0.12 <
\lambda_\gamma < 0.13$ where in the SM coupling parameters
$\kappa_\gamma = 1$ and $\lambda_\gamma = 0$ (for this analysis form factor $\Lambda =
2$~TeV). These limits are the best at a hadron collider and are a
direct examination of the $WW\gamma$ vertex.

\begin{figure}
\includegraphics[height=2.15in]{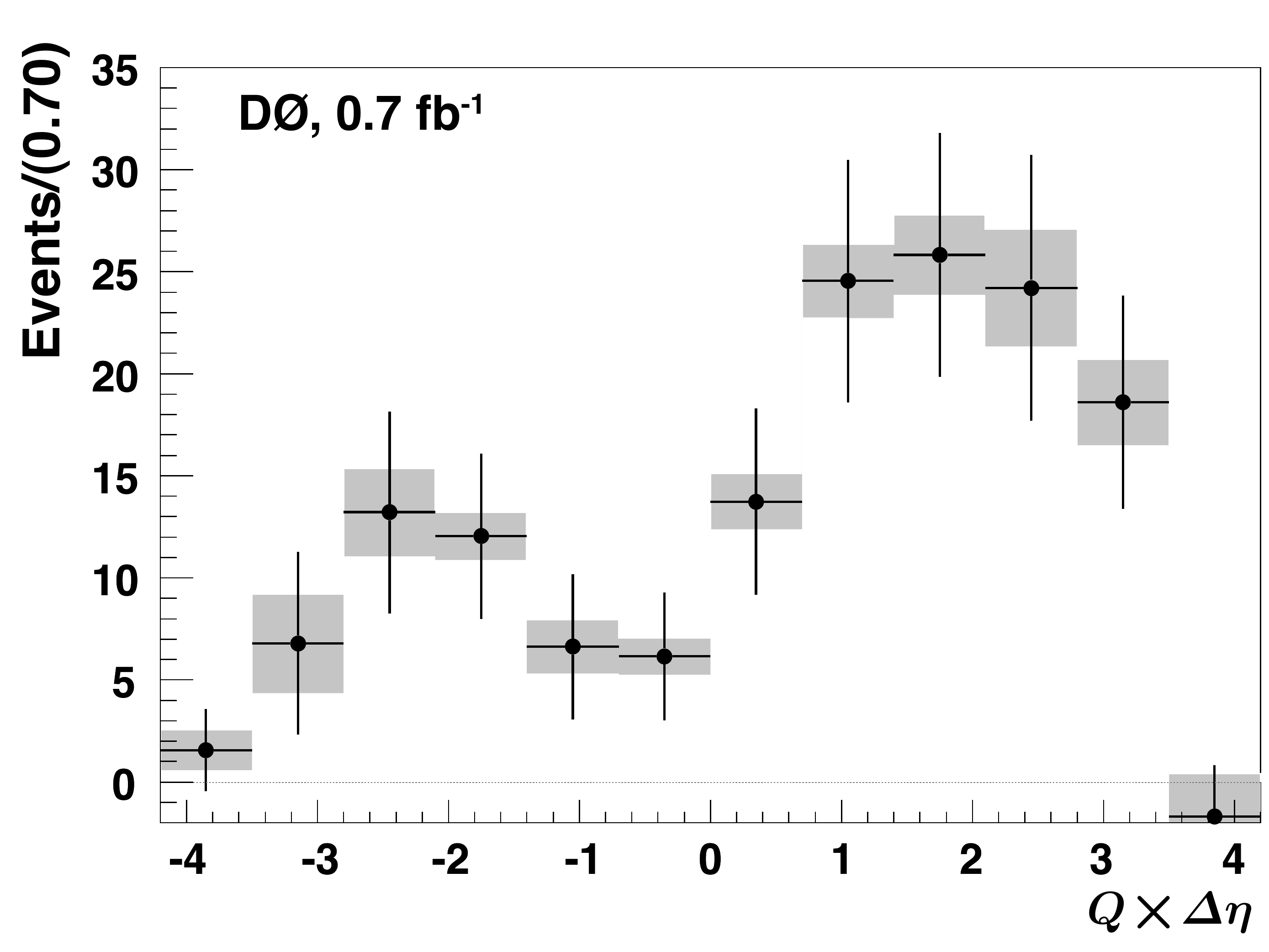}
\includegraphics[height=2.05in]{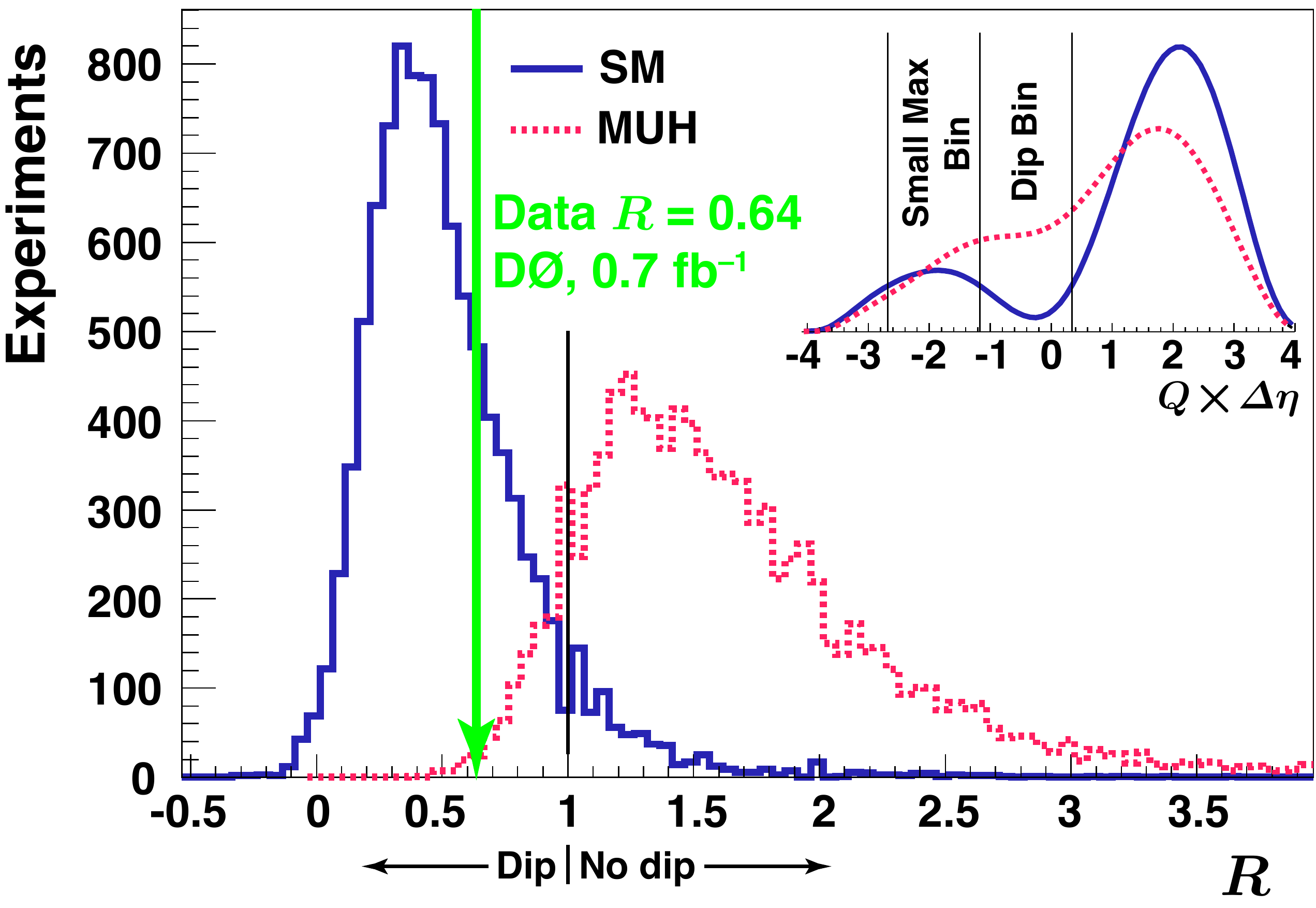}
\caption{\label{fig:csrd}  $Q_\ell \times
\Delta \eta$ distribution (left) and its analysis (right). }
\end{figure}

The background subtracted data are used to construct the $Q_\ell
\times \Delta \eta$ distribution as shown on the left in
Fig~\ref{fig:csrd}. Depicted on the right in Fig~\ref{fig:csrd} is a
study to determine the significance of the dip that is apparent in the
$Q_\ell \times \Delta \eta$ distribution. Two bins are chosen, one
that for the SM covers the dip region and an equally sized bin to its
left that samples the preceding peak (see the inset in the
figure). The ratio of events in the dip bin to the small maximum bin
($R$) is measured. If $R < 1$ then a dip is observed. A value of $R =
0.64$ is observed in the data. A minimal unimodal hypothesis (MUH) is
generated with anomalous couplings -- this model is on the verge of not
having a dip. 10,000 pseudo-experiments (SM and MUH) are performed to
determine the significance of the $R$ value. In the ensemble tests,
28\% of SM experiments had a higher value of $R$, indicating that the
measurement is consistent with the SM. Only 45 MUH experiments out of
10,000 had an $R$ value below the data. The conclusion is that the
probability that a MUH hypothesis could fluctuate to the data $R$
value or lower is $45/10,000; p = (4.5\pm 0.7)\times 10^{-3}$
corresponding to a Gaussian $2.6\sigma$. D0 thus makes the first
measurement of the $Q_\ell \times \Delta \eta$ distribution and it is
indicative of the RAZ of the Standard Model.

\section{CONCLUSIONS}
Diboson physics is extremely important for testing the Standard
Model. $W\gamma$ and $Z\gamma$ production show no hints of a
deviation from the SM. Therefore, limits were set on anomalous
couplings. The Radiation Amplitude Zero in the $W\gamma$ system is also
measured via the charge-signed rapidity distribution for the first
time. 

\begin{acknowledgements}

The author wishes to thank the staffs at Fermilab and collaborating
institutions as well as the respective funding agencies for CDF and
D0. 
\end{acknowledgements}

\end{document}